\documentclass[12pt,english]{article}
\usepackage{bm,bbm,lscape}
\usepackage[T1]{fontenc}
\usepackage{natbib, amsmath,authblk}
\usepackage{marvosym}
\usepackage{multicol}
\usepackage[british]{babel}
\usepackage{graphicx,xcolor,geometry,setspace,lineno}
\usepackage{pstricks}
\usepackage[british]{babel}
\usepackage[justification=justified,singlelinecheck=false,labelfont=it]{caption}
\usepackage{graphicx}
\usepackage{fancyhdr}
\definecolor{Mygrey}{gray}{0.65}

\title{Markov-switching generalized additive models}

\author{
  Roland Langrock\footnote{Corresponding author. E-mail: \text{roland.langrock@st-andrews.ac.uk}}\\
  \vspace{-1.4em}
  \textit{University of St Andrews}
  \and
    \vspace{-1em}
  Thomas Kneib\\
  \vspace{-1.4em}
  \textit{University of G\"ottingen}
  \and
    \vspace{-1em}
  Richard Glennie\\
  \vspace{-1.4em}
  \textit{University of St Andrews}
  \and
    \vspace{-1em}
  Th\'eo Michelot\\
  \vspace{-1.4em}
  \textit{INSA de Rouen}
    \vspace{-1.5em}
}

\date{}

\begin{document}

\begin{spacing}{1.4}
\maketitle


\begin{abstract}
\noindent
We consider Markov-switching regression models, i.e.\ models for time series regression analyses where the functional relationship between covariates and response is subject to regime switching controlled by an unobservable Markov chain. Building on the powerful hidden Markov model machinery and the methods for penalized B-splines routinely used in regression analyses, we develop a framework for nonparametrically estimating the functional form of the effect of the covariates in such a regression model, assuming an additive structure of the predictor. The resulting class of Markov-switching generalized additive models is immensely flexible, and contains as special cases the common parametric Markov-switching regression models and also generalized additive and generalized linear models. The feasibility of the suggested maximum penalized likelihood approach is demonstrated by simulation and further illustrated by modelling how energy price in Spain depends on the Euro/Dollar exchange rate.
\end{abstract}

\vspace{0.5em}

\vspace{0.5em}

\section{Introduction}\label{intro}

In regression scenarios where the data have a time series structure, there is often parameter instability with respect to time \citep{kim08}.
A popular strategy to account for such dynamic patterns is to employ regime switching where parameters vary in time, taking on finitely many values, controlled by an unobservable Markov chain.
Such models are referred to as Markov-switching or regime-switching regression models, following the seminal papers by \citet{gol73} and \citet{ham89}. A basic Markov-switching regression model involves a time series $\{Y_t\}_{t=1,\ldots,T}$ and an associated sequence of covariates $x_1,\ldots,x_T$ (including the possibility of $x_t=y_{t-1}$), with the relation between $x_t$ and $Y_t$ specified as
 \begin{linenomath*}
\begin{equation}\label{simplemodel}
Y_t = f^{(s_t)}(x_t) + \sigma_{s_t} \epsilon_t \, ,
\end{equation}
 \end{linenomath*}
where typically $\epsilon_t \stackrel{iid}{\sim} \mathcal{N} (0,1)$ and $s_t$ is the state at time $t$ of an unobservable $N$-state Markov chain. In other words, the functional form of the relation between $x_t$ and $Y_t$ and the residual variance change over time according to state switches of an underlying Markov chain, i.e.\ each state corresponds to a regime with different stochastic dynamics.
The Markov chain induces serial dependence, typically such that the states are persistent in the sense that regimes are active for longer periods of time, on average, than they would be if an independent mixture model was used to select among regimes. The classic example is an economic time series where the effect of an explanatory variable may differ between times of high and low economic growth \citep{ham08}. 

The simple model given in (\ref{simplemodel}) can be (and has been) modified in various ways, for example allowing for multiple covariates or for general error distributions from the generalized linear model (GLM) framework. An example for the latter is the Markov-switching Poisson regression model discussed in \citet{wan01}. However, in the existing literature the relationship between the target variable and the covariates is commonly specified in parametric form and usually assumed to be linear, with little investigation, if any, into the absolute or relative goodness of fit. The aim of the present work is to provide effective and accessible methods for a nonparametric estimation of the functional form of the predictor. These build on a) the strengths of the hidden Markov model (HMM) machinery \citep{zuc09}, in particular the forward algorithm, which allows for a simple and fast evaluation of the likelihood of a Markov-switching regression model (parametric or nonparametric), and b) the general advantages of penalized B-splines, i.e.\ P-splines \citep{eil96}, which we employ to obtain almost arbitrarily flexible functional estimators of the relationship between target variable and covariate(s). Model fitting is done via a numerical maximum penalized likelihood estimation, using either generalized cross-validation or an information criterion approach to select smoothing parameters that control the balance between goodness-of-fit and smoothness. Since parametric polynomial models are included as limiting cases for very large smoothing parameters, such a procedure also comprises the possibility to effectively reduce the functional effects to their parametric limiting cases, so that the conventional parametric Markov-switching regression models effectively are nested special cases of our more flexible models.

Our approach is by no means limited to models of the form given in (\ref{simplemodel}). In fact, the flexibility of the HMM machinery allows for the consideration of models from a much bigger class, which we term {\em Markov-switching generalized additive models} (MS-GAMs). These are simply generalized additive models (GAMs) with an additional time component, where the predictor --- including additive smooth functions of covariates, parametric terms and error terms --- is subject to regime changes controlled by an underlying Markov chain, analogously to (\ref{simplemodel}). While the methods do not necessitate a restriction to additive structures, we believe these to be most relevant in practice and hence have decided to focus on these models in the present work. 
Our work is closely related to that of \citet{des13}. Those authors, however, confine their consideration to the case of only one covariate and the identity link function.
Furthermore, we note that our approach is similar in spirit to that proposed in \citet{lan15}, where the aim is to nonparametrically estimate the densities of the state-dependent distributions of an HMM.


The paper is structured as follows. In Section \ref{MSGAMs}, we formulate general Markov-switching regression models, describe how to efficiently evaluate their likelihood, and develop the spline-based nonparametric estimation of the functional form of the predictor. The performance of the suggested approach is then investigated in three simulation experiments in Section \ref{simul}. In Section \ref{appl}, we apply the approach to Spanish energy price data to demonstrate its feasibility and potential. We conclude in Section \ref{discuss}.

\section{Markov-switching generalized additive models}
\label{MSGAMs}

\subsection{Markov-switching regression models}

We begin by formulating a Markov-switching regression model with arbitrary form of the predictor, encompassing both parametric and nonparametric specifications. Let $\{Y_t\}_{t=1,\ldots,T}$ denote the target variable of interest (a time series), and let $x_{p1},\ldots,x_{pT}$ denote the associated values of the $p$-th covariate considered, where $p=1,\ldots,P$. We summarize the covariate values at time $t$ in the vector $\mathbf{x}_{\cdot t}=(x_{1t},\ldots,x_{Pt})$. Further let $s_1,\ldots,s_T$ denote the states of an underlying unobservable $N$-state Markov chain $\{S_t\}_{t=1,\ldots,T}$. Finally, we assume that conditional on $(s_t,\mathbf{x}_{\cdot t})$, $Y_t$ follows some distribution from the exponential family and is independent of all other states, covariates and observations. We write
 \begin{linenomath*}
\begin{equation}\label{expform}
 g\bigl( \mathbbm{E}(Y_t \mid s_t,\mathbf{x}_{\cdot t}) \bigr) = \eta^{(s_t)}(\mathbf{x}_{\cdot t}),
\end{equation} 
 \end{linenomath*}
where $g$ is some link function, typically the canonical link function associated with the exponential family distribution considered. That is, the expectation of $Y_t$ is linked to the covariate vector $\mathbf{x}_{\cdot t}$ via the predictor function $\eta^{(i)}$, which maps the covariate vector to $\mathbbm{R}$, when the underlying Markov chain is in state $i$, i.e.\ $S_t=i$. Essentially there is one regression model for each state $i$, $i=1,\ldots,N$. In the following, we use the shorthand $\mu_t^{(s_t)}=\mathbbm{E}(Y_t \mid s_t,\mathbf{x}_{\cdot t})$.

To fully specify the conditional distribution of $Y_t$, additional parameters may be required, depending on the error distribution considered. For example, if $Y_t$ is conditionally Poisson distributed, then (\ref{expform}) fully specifies the state-dependent distribution (e.g.\ with $g(\mu)=\log(\mu)$), whereas if $Y_t$ is normally distributed (in which case $g$ usually is the identity link), then the variance of the error needs to be specified, and would typically be assumed to also depend on the current state of the Markov chain. We use the notation $\phi^{(s_t)}$ to denote such additional time-dependent parameters (typically dispersion parameters), and denote the conditional density of $Y_t$, given $(s_t,\mathbf{x}_{\cdot t})$, as $p_Y(y_t ,\mu_t^{(s_t)},\phi^{(s_t)})$. The simplest and probably most popular such model assumes a conditional normal distribution for $Y_t$, a linear form of the predictor and a state-dependent error variance, leading to the model
 \begin{linenomath*}
\begin{equation}\label{linpred}
Y_t = \beta_{0}^{(s_t)} + \beta_1^{(s_t)} x_{1t} + \ldots + \beta_p^{(s_t)} x_{Pt} + \sigma_{s_t}\epsilon_t,
\end{equation}
 \end{linenomath*}
where $\epsilon_t \stackrel{iid}{\sim} \mathcal{N}(0,1)$ (cf.\ \citealp{fru06}; \citealp{kim08}).

Assuming homogeneity of the Markov chain --- which can easily be relaxed if desired --- we summarize the probabilities of transitions between the different states in the $N \times N$ transition probability matrix (t.p.m.) $\boldsymbol{\Gamma}=\left( \gamma_{ij} \right)$, where $\gamma_{ij}=\Pr \bigl(S_{t+1}=j\vert S_t=i \bigr)$, $i,j=1,\ldots,N$. The initial state probabilities are summarized in the row vector $\boldsymbol{\delta}$, where $\delta_{i} = \Pr (S_1=i)$, $i=1,\ldots,N$. It is usually convenient to assume $\boldsymbol{\delta}$ to be the stationary distribution, which, if it exists, is the solution to $\boldsymbol{\delta}\boldsymbol{\Gamma}=\boldsymbol{\delta}$ subject to $\sum_{i=1}^N \delta_i=1$.

\subsection{Likelihood evaluation by forward recursion}

A Markov-switching regression model, with conditional density $p_Y(y_t,\mu_t^{(s_t)},\phi^{(s_t)})$ and underlying Markov chain characterized by $(\boldsymbol{\Gamma},\boldsymbol{\delta})$, can be regarded as an HMM with additional dependence structure (here in the form of covariate influence); see \citet{zuc09}. This opens up the way for exploiting the efficient and flexible HMM machinery. Most importantly, irrespective of the type of exponential family distribution considered, an extremely efficient recursion can be applied in order to evaluate the likelihood of a Markov-switching regression model, namely the so-called {forward algorithm}. To see this, consider the vectors of forward variables, defined as the row vectors
 \begin{linenomath*}
\begin{align*}
\boldsymbol{\alpha}_t & = \bigl( {\alpha}_t (1), \ldots , {\alpha}_t (N) \bigr) \, , \; t=1,\ldots,T, \\
\text{where } \; {\alpha}_t (j) & = p (y_1, \ldots, y_t, S_t=j \mid \mathbf{x}_{\cdot 1} \ldots \mathbf{x}_{\cdot t}) \; \text{ for } \; j=1,\ldots,N \, .
\end{align*}
 \end{linenomath*}
Here $p$ is used as a generic symbol for a (joint) density. Then the following recursive scheme can be applied:
 \begin{linenomath*}
 \begin{align}\label{forw}
 \nonumber \boldsymbol{\alpha}_1 & = \boldsymbol{\delta} \mathbf{Q}(y_1) \, , \\
 \boldsymbol{\alpha}_{t} & =  \boldsymbol{\alpha}_{t-1} \boldsymbol{\Gamma} \mathbf{Q}(y_{t}) \quad (t=2,\ldots,T)\, ,
 \end{align}
 \end{linenomath*}
where $$\mathbf{Q}(y_t)= \text{diag} \bigl( p_Y(y_t,\mu_t^{(1)},\phi^{(1)}), \ldots, p_Y(y_t,\mu_t^{(N)},\phi^{(N)}) \big).$$
The recursion (\ref{forw}) follows immediately from $${\alpha}_{t} (j)= \sum_{i=1}^N {\alpha}_{t-1}(i) \gamma_{ij} p_Y(y_t,\mu_t^{(j)},\phi^{(j)}),$$ which in turn can be derived in a straightforward manner using the model's dependence structure (for example analogously to the proof of Proposition 2 in \citealp{zuc09}). 
The likelihood can thus be written as a matrix product:
 \begin{linenomath*}
\begin{equation}\label{lik}
\mathcal{L}(\boldsymbol{\theta}) = \sum_{i=1}^N {\alpha}_T(i) = \boldsymbol{\delta} \mathbf{Q}(y_1) \boldsymbol{\Gamma} \mathbf{Q}(y_{2}) \ldots \boldsymbol{\Gamma} \mathbf{Q}(y_{T}) \mathbf{1} \, ,
\end{equation}
 \end{linenomath*}
where $\mathbf{1}\in \mathbbm{R}^N$ is a column vector of ones, and where $\boldsymbol{\theta}$ is a vector comprising all model parameters. The computational cost of evaluating (\ref{lik}) is {linear} in the number of observations, $T$, such that a numerical maximization of the likelihood is feasible in most cases, even for very large $T$ and moderate numbers of states $N$.

\subsection{Nonparametric modelling of the predictor}\label{nonp}

Notably, the likelihood form given in (\ref{lik}) applies for any form of the conditional density $p_Y(y_t,\mu_t^{(s_t)},\phi^{(s_t)})$. In particular, it can be used to estimate simple Markov-switching regression models, e.g.\ with linear predictors, or in fact with any GLM-type structure within states. Here we are concerned with a nonparametric estimation of the functional relationship between $Y_t$ and $\mathbf{x}_{\cdot t}$. To achieve this, we consider a GAM-type framework \citep{woo06}, with the predictor comprising additive smooth state-dependent functions of the covariates:
 \begin{linenomath*}
\begin{equation*}
 g( \mu_t^{(s_t)} ) =  \eta^{(s_t)}(\mathbf{x}_{\cdot t}) = \beta_0^{(s_t)} + f_1^{(s_t)}(x_{1t}) + f_2^{(s_t)}(x_{2t}) + \ldots + f_P^{(s_t)}(x_{Pt}).
\end{equation*}
 \end{linenomath*}
We simply have one GAM associated with each state of the Markov chain.
To achieve a flexible estimation of the functional form, we express each of the functions $f_p^{(i)}$, $i=1,\ldots,N$, $p=1,\ldots,P$, as a finite linear combination of a high number of basis functions, $B_{1}, \ldots , B_{K}$:
 \begin{linenomath*}
\begin{equation}\label{lincom}
	f_p^{(i)}(x) = \sum_{k=1}^K \gamma_{ipk} B_k(x).
\end{equation}
 \end{linenomath*}
Note that different sets of basis functions can be applied to represent the different functions, but to keep the notation simple we here consider a common set of basis functions for all $f_p^{(i)}$. A common choice for the basis is to use B-splines, which form a numerically stable, convenient basis for the space of polynomial splines, i.e.\ piecewise polynomials that are fused together smoothly at the interval boundaries; see \citet{deb78} and \citet{eil96} for more details. We use cubic B-splines, in ascending order in the basis used in (\ref{lincom}). The number of B-splines considered, $K$, determines the flexibility of the functional form, as an increasing number of basis functions allows for an increasing curvature of the function being modeled. Instead of trying to select an optimal number of basis elements, we follow \citet{eil96} and modify the likelihood by including a difference penalty on coefficients of adjacent B-splines. The number of basis B-splines, $K$, then simply needs to be sufficiently large in order to yield high flexibility for the functional estimates. Once this threshold is reached, a further increase in the number of basis elements no longer changes the fit to the data due to the impact of the penalty. Considering second-order differences --- which leads to an approximation of the integrated squared curvature of the function estimate \citep{eil96} --- leads to the difference penalty $0.5 \lambda_{ip} \sum_{k=3}^K (\Delta^2 \gamma_{ipk})^2$,
where $\lambda_{ip} \geq 0$ are smoothing parameters and where $\Delta^2 \gamma_{ipk} = \gamma_{ipk}-2\gamma_{ip,k-1}+\gamma_{ip,k-2}$.

We then modify the (log-)likelihood of the MS-GAM --- specified by $p_Y(y_t,\mu_t^{(s_t)},\phi^{(s_t)})$ in combination with (\ref{lincom}) and underlying Markov chain characterized by $(\boldsymbol{\Gamma},\boldsymbol{\delta})$ --- by including the above difference penalty, one for each of the smooth functions appearing in the state-dependent predictors:
 \begin{linenomath*}
\begin{equation}\label{penlik}
l_{\text{pen.}} = \log \bigl( \mathcal{L} (\boldsymbol{\theta}) \bigr) - \sum_{i=1}^N \sum_{p=1}^P \dfrac{\lambda_{ip}}{2} \sum_{k=3}^K (\Delta^2 \gamma_{ipk})^2\ .
\end{equation}
 \end{linenomath*}

The maximum penalized likelihood estimate then reflects a compromise between goodness-of-fit and smoothness, where an increase in the smoothing parameters leads to an increased emphasis being put on smoothness. We discuss the choice of the smoothing parameters in more detail in the subsequent section. As $\lambda_{ip} \rightarrow\infty$, the corresponding penalty dominates the log-likelihood, leading to a sequence of estimated coefficients $\gamma_{ip1}, \ldots , \gamma_{ipK}$ that are on a straight line. Thus, we obtain the common linear predictors, as given in (\ref{linpred}), as a limiting case. Similarly, we can obtain parametric functions with arbitrary polynomial order $q$ as limiting cases by considering $(q+1)$-th order differences in the penalty. The common parametric regression models thus are essentially nested within the class of nonparametric models that we consider. One can of course obtain these nested special cases more directly, by simply specifying parametric rather than nonparametric forms for the predictor. On the other hand, it can clearly be advantageous not to constrain the functional form in any way {\em a priori}, though still allowing for the possibility of obtaining constrained parametric cases as a result of a data-driven choice of the smoothing parameters. Standard GAMs and even GLMs are also nested in the considered class of models ($N=1$), but this observation is clearly less relevant, since powerful software is already available for these special cases.

\subsection{Inference}
\label{estim}

For given smoothing parameters, all model parameters --- including the parameters determining the Markov chain, any dispersion parameters, the coefficients $\gamma_{ipk}$ used in the linear combinations of B-splines and any other parameters required to specify the predictor --- can be estimated simultaneously by numerically maximizing the penalized log-likelihood given in (\ref{penlik}). For each function $f_p^{(i)}$, $i=1,\ldots,N$, $p=1,\ldots,P$, one of the coefficients needs to be fixed to render the model identifiable, such that the intercept controls the height of the predictor function. A convenient strategy to achieve this is to first standardize each sequence of covariates $x_{p1},\ldots,x_{pT}$, $p=1,\ldots,P$, shifting all values by the sequence's mean and dividing the shifted values by the sequence's standard deviation, and second consider an odd number of B-spline basis functions $K$ with $\gamma_{ip,(K+1)/2}=0$ fixed.
The numerical maximization is carried out subject to well-known technical issues arising in all optimization problems, including parameter constraints and local maxima of the likelihood. The latter can be either easy to deal with or a challenging problem, depending on the complexity of the model considered. 

Uncertainty quantification, on both the estimates of parametric parts of the model and on the function estimates, can be performed based on the approximate covariance matrix available as the inverse of the observed Fisher information or using a parametric bootstrap \citep{efr93}. The latter avoids relying on asymptotics, which is particularly problematic when the number of B-spline basis functions increases with the sample size. From the bootstrap samples, we can obtain pointwise as well as simultaneous confidence intervals for the estimated regression functions. Pointwise confidence intervals are simply given via appropriate quantiles obtained from the bootstrap replications. Simultaneous confidence bands are obtained by scaling the pointwise confidence intervals until they contain a pre-specified fraction of all bootstrapped curves completely \citep{kri10}.

For the closely related class of parametric HMMs, identifiability holds under fairly weak conditions, which in practice will usually be satisfied, namely that the t.p.m.\ of the unobserved Markov chain has full rank and that the state-specific distributions are distinct \citep{gas13}. This result transfers to the more general class of MS-GAMs if, additionally, the state-specific GAMs are identifiable. Conditions for the latter are simply the same as in any standard GAM and in particular the nonparametric functions have to be centered around zero. Furthermore, in order to guarantee estimability of a smooth function on a given domain, it is necessary that the covariate values cover that domain sufficiently well. In practice, i.e.\ when dealing with finite sample sizes, parameter estimation will be difficult if the level of correlation, as induced by the unobserved Markov chain, is low, and also if the state-specific GAMs are similar. The stronger the correlation in the state process, the clearer becomes the pattern and hence the easier it is for the model to allocate observations to states. Similarly, the estimation performance will be best, in terms of numerical stability, if the state-specific GAMs are clearly distinct. (See also the simulation experiments in Section \ref{simul} below.)

\subsection{Choice of the smoothing parameters}
\label{smoothies}

In Section~\ref{estim}, we described how to fit an MS-GAM to data for a {\em given} smoothing parameter vector. To choose adequate smoothing parameters in a data-driven way, generalized cross-validation can be applied. 
A leave-one-out cross-validation will typically be computationally infeasible. Instead, for a given time series to be analyzed, we generate $C$ random partitions such that in each partition a high percentage of the observations, e.g.\ 90\%, form the calibration sample, while the remaining observations constitute the validation sample. For each of the $C$ partitions and any  $\boldsymbol{\lambda}=(\lambda_{11},\ldots,\lambda_{1P},\ldots,\lambda_{N1},\ldots,\lambda_{NP})$, the model is then calibrated by estimating the parameters using only the calibration sample (treating the data points from the validation sample as missing data, which is straightforward using the HMM forward algorithm; see \citealp{zuc09}). Subsequently, proper scoring rules \citep{gne07} can be used on the validation sample to assess the model for the given $\boldsymbol{\lambda}$ and the corresponding calibrated model. For computational convenience, we consider the log-likelihood of the validation sample, under the model fitted in the calibration stage, as the score of interest (now treating the data points from the calibration sample as missing data). From some pre-specified grid $\boldsymbol{\Lambda} \subset \mathbbm{R}_{\geq 0}^{N\times P}$, we then select the $\boldsymbol{\lambda}$ that yields the highest mean score over the $C$ cross-validation samples. The number of samples $C$ needs to be high enough to give meaningful scores (i.e.\ such that the scores give a clear pattern rather than noise only; from our experience, $C$ should not be smaller than 10), but must not be too high to allow for the approach to be computationally feasible.

An alternative, less computer-intensive approach for selecting the smoothing parameters is based on the Akaike Information Criterion (AIC), calculating, for each smoothing parameter vector from the grid considered, the following  AIC-type statistic:
$$ \text{AIC}_p = -2 \log \mathcal{L} + 2 \nu.  $$
Here $\mathcal{L}$ is the unpenalized likelihood under the given model (fitted via penalized maximum likelihood), and $\nu$ denotes the effective degrees of freedom, defined as the trace of the product of the Fisher information matrix for the unpenalized likelihood and the inverse Fisher information matrix for the penalized likelihood \citep{gra92}. Using the effective degrees of freedom accounts for the effective dimensionality reduction of the parameter space resulting from the penalization. From all smoothing parameter vectors considered, the one with the smallest $\text{AIC}_p$ value is chosen.


\section{Simulation experiments}\label{simul}

\begin{figure}[tbh]
\centering
\includegraphics[width=0.76\textwidth]{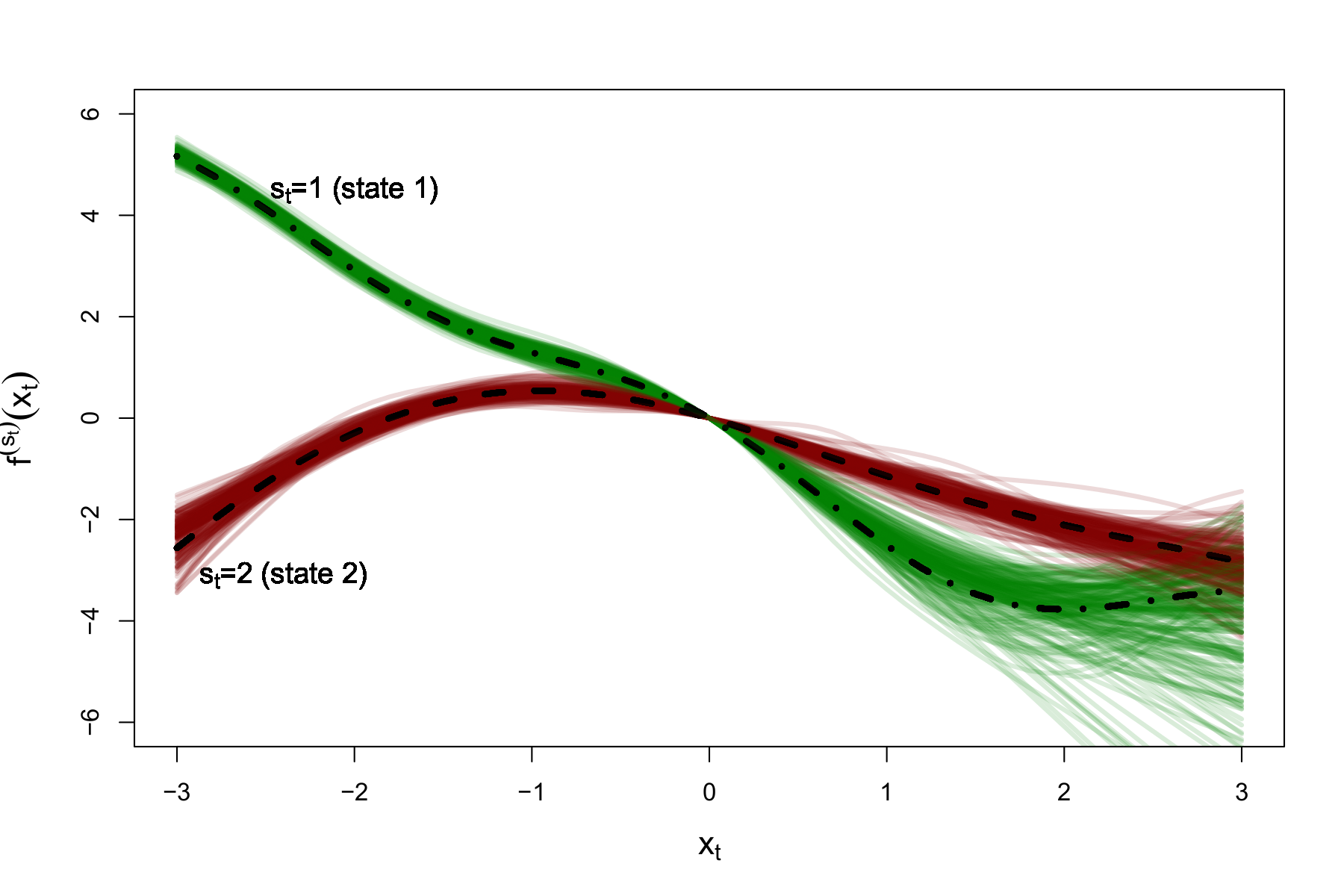}
\vspace{-0.5em}
\caption{Displayed are the true functions $f^{(1)}$ and $f^{(2)}$ used in {\em Scenario I} (dashed lines) and their estimates obtained in 200 simulation runs (green and red lines for states 1 and 2, respectively).}\label{fig1}
\end{figure}

{\em Scenario I.} We first consider a relatively simple scenario, with a Poisson-distributed target variable, a 2-state Markov chain selecting the regimes and only one covariate:
$$Y_t \sim \text{Poisson} (\mu_t^{(s_t)}) \, ,$$
where
$$\log ( \mu_t^{(s_t)} ) = \beta_0^{(s_t)} + f^{(s_t)}(x_{t}) \, .$$
The functional forms chosen for $f^{(1)}$ and $f^{(2)}$ are displayed by the dashed curves in Figure \ref{fig1}; both functions go through the origin. We further set $\beta_0^{(1)}=\beta_0^{(2)}=2$ and
$$\boldsymbol{\Gamma}=\begin{pmatrix} 0.9 & 0.1 \\ 0.1 & 0.9 \end{pmatrix} \, .$$
All covariate values were drawn independently from a uniform distribution on $[-3,3]$. We ran 200 simulations, in each run generating $T=300$ observations from the model described. An MS-GAM, with Poisson-distributed response and log-link, was then fitted via numerical maximum penalized likelihood estimation, as described in Section \ref{estim} above, using the optimizer \texttt{nlm} in R. We set $K=15$, hence using 15 B-spline basis densities in the representation of each functional estimate.

We implemented both generalized cross-validation and the AIC-based approach for choosing the smoothing parameter vector from a grid $\boldsymbol{\Lambda}= \Lambda_1 \times \Lambda_2$, where $\Lambda_1=\Lambda_2=\{0.125,1,8,64,512,4096\}$. We considered $C=25$ folds in the cross-validation. For both approaches, we estimated the mean integrated squared error (MISE) for the two functional estimators, as follows:
$$ \widehat{\text{MISE}}_{f^{(j)}} = \frac{1}{200}\sum_{z=1}^{200} \left( \int_{-3}^{3} \biggl( \hat{f}_z^{(j)}(x) - {f}^{(j)}(x) \biggr)^2 dx \right)  ,$$
for $j=1,2$, where $\hat{f}_z^{(j)}(x)$ is the functional estimate of ${f}^{(j)}(x)$ obtained in simulation run $z$. Using cross-validation, we obtained $\widehat{\text{MISE}}_{f^{(1)}}=1.808$ and $\widehat{\text{MISE}}_{f^{(2)}}=0.243$, while using the AIC-type criterion we obtained the slightly better values $\widehat{\text{MISE}}_{f^{(1)}}=1.408$ and $\widehat{\text{MISE}}_{f^{(2)}}=0.239$. In the following, we report the results obtained using the AIC-based approach.

The sample mean estimates of the transition probabilities $\gamma_{11}$ and $\gamma_{22}$ were obtained as $0.894$ (Monte Carlo standard deviation of estimates: $0.029$) and $0.896$ ($0.032$), respectively. The estimated functions $\hat{f}^{(1)}$ and $\hat{f}^{(2)}$ from all 200 simulation runs are visualized in Figure \ref{fig1}. The functions have been shifted so that they go through the origin. All fits are fairly reasonable. The sample mean estimates of the predictor value for $x_{t}=0$ were obtained as $2.002$ ($0.094$) and $1.966$ ($0.095$) for states 1 and 2, respectively. \\

\begin{figure}[tbh]
\centering
\includegraphics[width=1\textwidth]{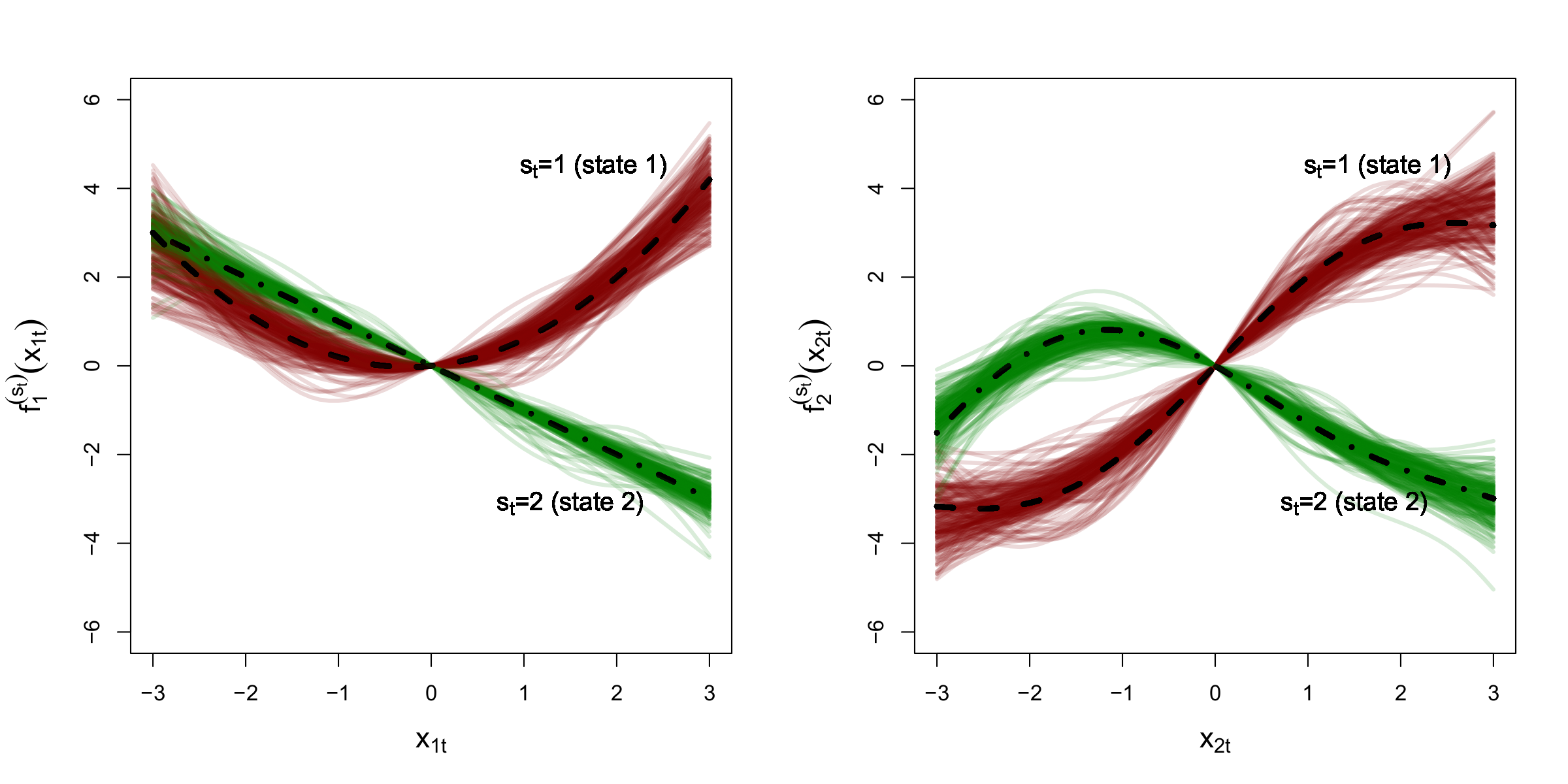}
\vspace{-0.5em}
\caption{Displayed are the true functions $f_1^{(1)}$, $f_1^{(2)}$, $f_2^{(1)}$ and $f_2^{(2)}$ used in {\em Scenario II} (dashed lines) and their estimates obtained in 200 simulation runs (red and green lines for states 1 and 2, respectively; $f_1^{(1)}$ and $f_1^{(2)}$, which describe the state-dependent effect of the covariate $x_{1t}$ on the predictor, and corresponding estimates are displayed in the left panel; $f_2^{(1)}$ and $f_2^{(2)}$, which describe the state-dependent effect of the covariate $x_{2t}$ on the predictor, and corresponding estimates are displayed in the right panel).}\label{fig2}
\end{figure}

\noindent
{\em Scenario II.} The second simulation experiment we conducted is slightly more involved, with a normally distributed target variable, an underlying 2-state Markov chain and now two covariates:
$$Y_t \sim \mathcal{N} (\mu_t^{(s_t)},\sigma_{s_t})\, ,$$
where
$$\mu_t^{(s_t)} = \beta_0^{(s_t)} + f_1^{(s_t)}(x_{1t})+ f_2^{(s_t)}(x_{2t}) \, .$$
The functional forms we chose for $f_1^{(1)}$, $f_1^{(2)}$, $f_2^{(1)}$ and $f_2^{(2)}$ are displayed in Figure \ref{fig2}; again all functions go through the origin.
We further set $\beta_0^{(1)}=1$, $\beta_0^{(2)}=-1$, $\sigma_{1}=3$, $\sigma_{2}=2$ and
$$\boldsymbol{\Gamma}=\begin{pmatrix} 0.95 & 0.05 \\ 0.05 & 0.95 \end{pmatrix} \, .$$
The covariate values were drawn independently from a uniform distribution on $[-3,3]$. In each of 200 simulation runs, $T=1000$ observations were generated.

For the choice of the smoothing parameter vector, we considered the grid $\boldsymbol{\Lambda}= \Lambda_1 \times \Lambda_2 \times \Lambda_3 \times \Lambda_4$, where $\Lambda_1=\Lambda_2=\Lambda_3=\Lambda_4=\{0.25,4,64,1024,16384\}$. The AIC-based smoothing parameter selection led to MISE estimates that overall were marginally lower than their counterparts obtained when using cross-validation ($0.555$ compared to $0.565$, averaged over all four functions being estimated), so again in the following we report the results obtained based on the AIC-type criterion. The (true) function $f_1^{(2)}$ is in fact a straight line, and, notably, the associated smoothing parameter was chosen as $16384$, hence as the maximum possible value from the grid considered, in 129 out of the 200 cases, whereas for example for the function $f_2^{(2)}$, which has a moderate curvature, the value $16384$ was not chosen even once as the smoothing parameter.

In this experiment, the sample mean estimates of the transition probabilities $\gamma_{11}$ and $\gamma_{22}$ were obtained as $0.950$ (Monte Carlo standard deviation of estimates: $0.011$) and $0.948$ ($0.012$), respectively. The estimated functions $\hat{f}_1^{(1)}$, $\hat{f}_1^{(2)}$, $\hat{f}_2^{(1)}$ and $\hat{f}_2^{(2)}$ from all 200 simulation runs are displayed in Figure \ref{fig2}. Again all have been shifted so that they go through the origin. The sample mean estimates of the predictor value for $x_{1t}=x_{2t}=0$ were $0.989$ ($0.369$) and $-0.940$ ($0.261$) for states 1 and 2, respectively. The sample mean estimates of the state-dependent error variances, $\sigma_{1}$ and $\sigma_{2}$, were obtained as $2.961$ ($0.107$) and $1.980$ ($0.078$), respectively. Again the results are very encouraging, with not a single simulation run leading to a complete failure in terms of capturing the overall pattern. \\ 

\begin{figure}[tbh]
\centering
\includegraphics[width=0.76\textwidth]{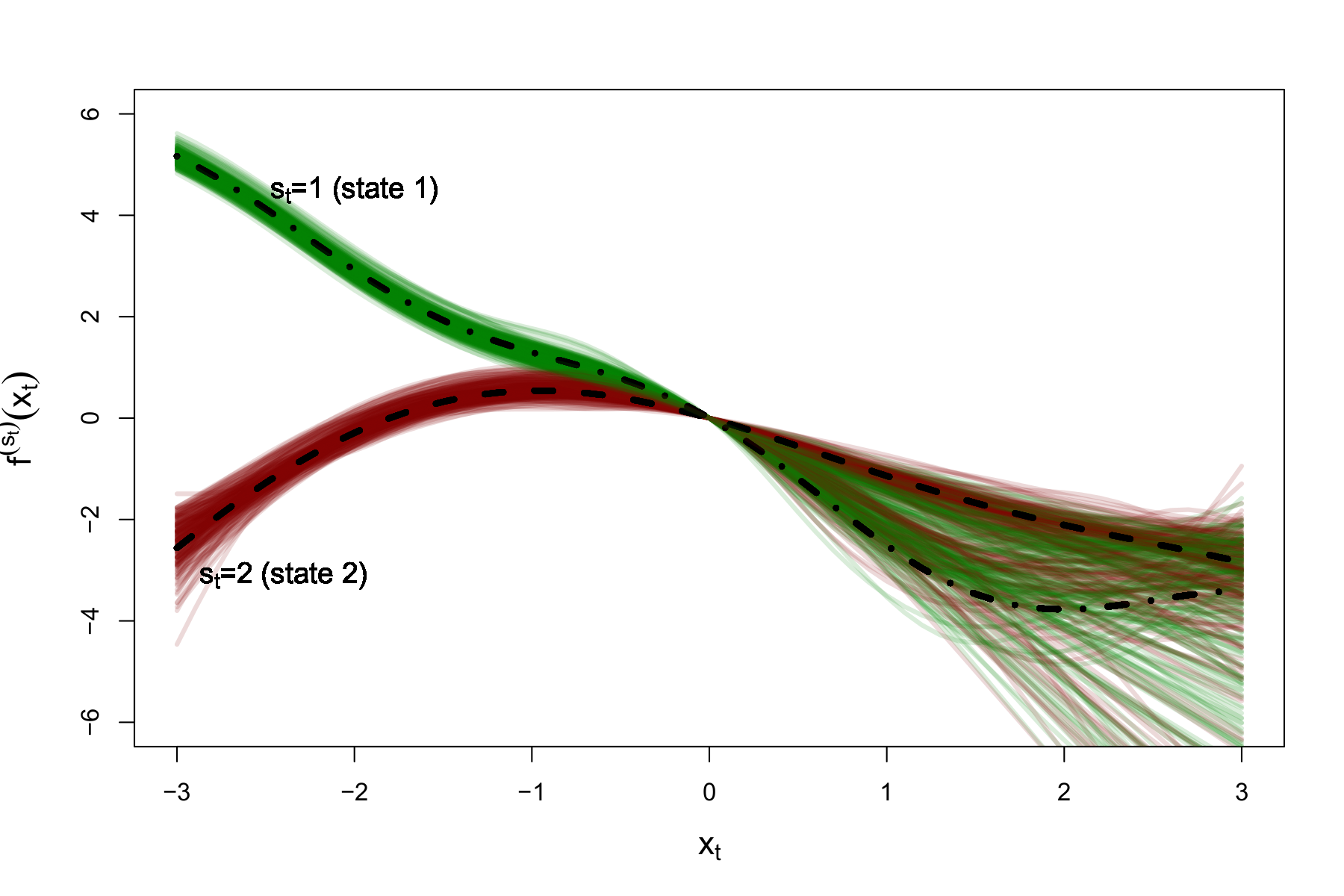}
\vspace{-0.5em}
\caption{Displayed are the true functions $f^{(1)}$ and $f^{(2)}$ used in {\em Scenario III} (dashed lines) and their estimates obtained in 200 simulation runs (green and red lines for states 1 and 2, respectively).}\label{fig3}
\end{figure}

\noindent
{\em Scenario III.} The estimator behavior both in {\em Scenario I} and in {\em Scenario II} is encouraging, and demonstrates that inference in MS-GAMs is clearly practicable in these two settings, both of which may occur in similar form in real data. However, as discussed in Section \ref{estim}, in some circumstances, parameter identification in finite samples can be difficult, especially if the level of correlation as induced by the Markov chain is low. To illustrate this, we re-ran {\em Scenario I}, using the exact same configuration as described above except that we changed $\boldsymbol{\Gamma}$ to
$$\boldsymbol{\Gamma}=\begin{pmatrix} 0.6 & 0.4 \\ 0.4 & 0.6 \end{pmatrix} \, .$$
In other words, compared to {\em Scenario I}, there is substantially less autocorrelation in the series that are generated. Figure \ref{fig3} displays the estimated functions $\hat{f}^{(1)}$ and $\hat{f}^{(2)}$ in this slightly modified scenario. Due to the fairly low level of autocorrelation, the estimator performance is substantially worse than in {\em Scenario I}, and in several simulation runs the model failed to capture the overall pattern, by allocating pairs $(y_t,x_t)$ with high values of the covariate $x_t$ to the wrong state of the Markov chain. The deterioration in the estimator performance is also reflected by higher standard errors: The sample mean estimates of the transition probabilities $\gamma_{11}$ and $\gamma_{22}$ were obtained as $0.590$ (Monte Carlo standard deviation of estimates: $0.082$) and $0.593$ ($0.088$), respectively, and the sample mean estimates of the predictor value for $x_{t}=0$ were obtained as $1.960$ ($0.151$) and $2.017$ ($0.145$) for states 1 and 2, respectively.

\section{Illustrating example: Spanish energy prices}\label{appl}

We analyze the data collected on the daily price of energy in Spain between 2002 and 2008. The data, 1784 observations in total, are available in the R package \texttt{MSwM} \citep{san14}. We consider the relationship over time between the price of energy $P_t$ and the Euro/Dollar exchange rate $r_t$. The stochastic volatility of financial time series makes the assumption of a fixed relationship between these two variables over time questionable, and it seems natural to consider a Markov-switching regime model. It is also probable that their relationship within a regime is non-linear with unknown functional form, thereby motivating the use of flexible nonparametric predictor functions. 
Here, our aim is to illustrate the potential benefits of considering flexible MS-GAMs rather than GAMs or parametric Markov-switching models when analyzing time series regression data.

To this end, we consider four different models. As benchmark models, we considered two parametric models with state-dependent linear predictor $\beta_0^{(s_t)} + \beta_1^{(s_t)} r_t$, with one (LIN) and two states (MS-LIN), respectively, assuming the response variable $P_t$ to be normally distributed. Additionally, we considered two nonparametric models as introduced in Section \ref{nonp}, with one state (hence a basic GAM) and two states (MS-GAM), respectively. In these two models, we assumed $P_t$ to be gamma-distributed, applying the log link function to meet the range restriction for the (positive) mean. 

\begin{figure}[htb]
  \centering
  \includegraphics[width=0.8\textwidth]{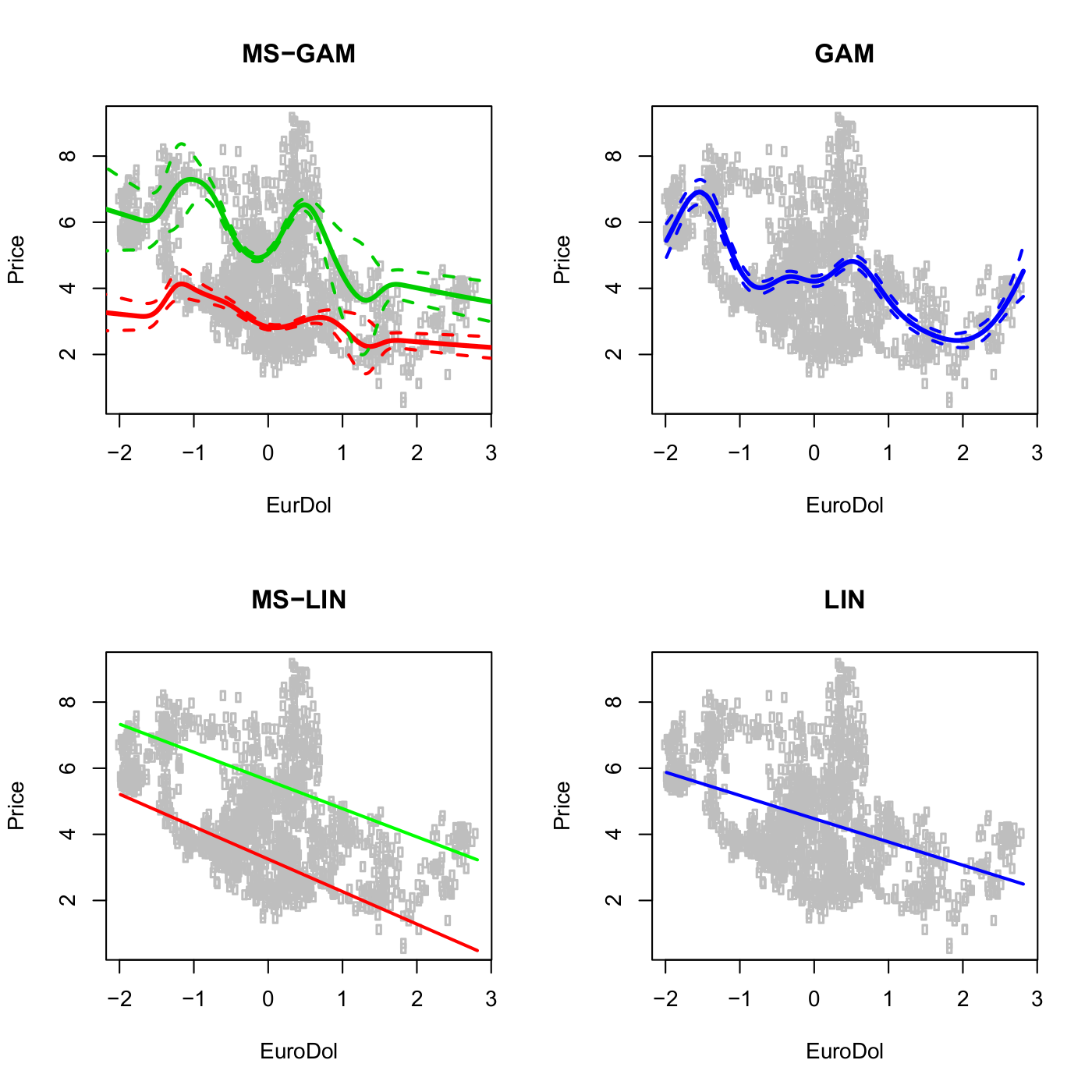}
  \caption{Observed energy price against Euro/Dollar exchange rate (gray points), with estimated state-dependent mean energy prices (solid lines) for one-state (blue) and two-state (green and red) nonparametric and linear models; nonparametric models are shown together with associated approximate 95\% pointwise confidence intervals obtained based on 999 parametric bootstrap samples (dotted lines).}
  \label{fig:real_models}
\end{figure}

Figure \ref{fig:real_models} shows the fitted curves for each model. For each one-state model (GAM and LIN), the mean curve passes through a region with no data (for values of $r_t$ around $-1$). This results in response residuals with clear systematic deviation. It is failings such as this which demonstrate the need for regime-switching models. Models were formally compared using one-step-ahead forecast evaluation, by means of the sum of the log-likelihoods of observations $P_u$ under the models fitted to all preceding observations, $P_1,\ldots,P_{u-1}$, considering $u=501,\ldots,1784$  (such that models are fitted to a reasonable number of observations). We obtained the following log-likelihood scores for each model: $-2314$ for LIN, $-2191$ for GAM, $-2069$ for MS-LIN and $-1703$ for MS-GAM. Thus, in terms of out-of-sample forecasts, the MS-GAM performed much better than any other model considered. Both two-state models performed much better than the single-state models, however the inflexibility of the MS-LIN model resulted in a poorer performance than that of its nonparametric counterpart, as clear non-linear features in the regression data are ignored. 



\begin{figure}[htb]
  \centering
  \includegraphics[width=0.8\textwidth]{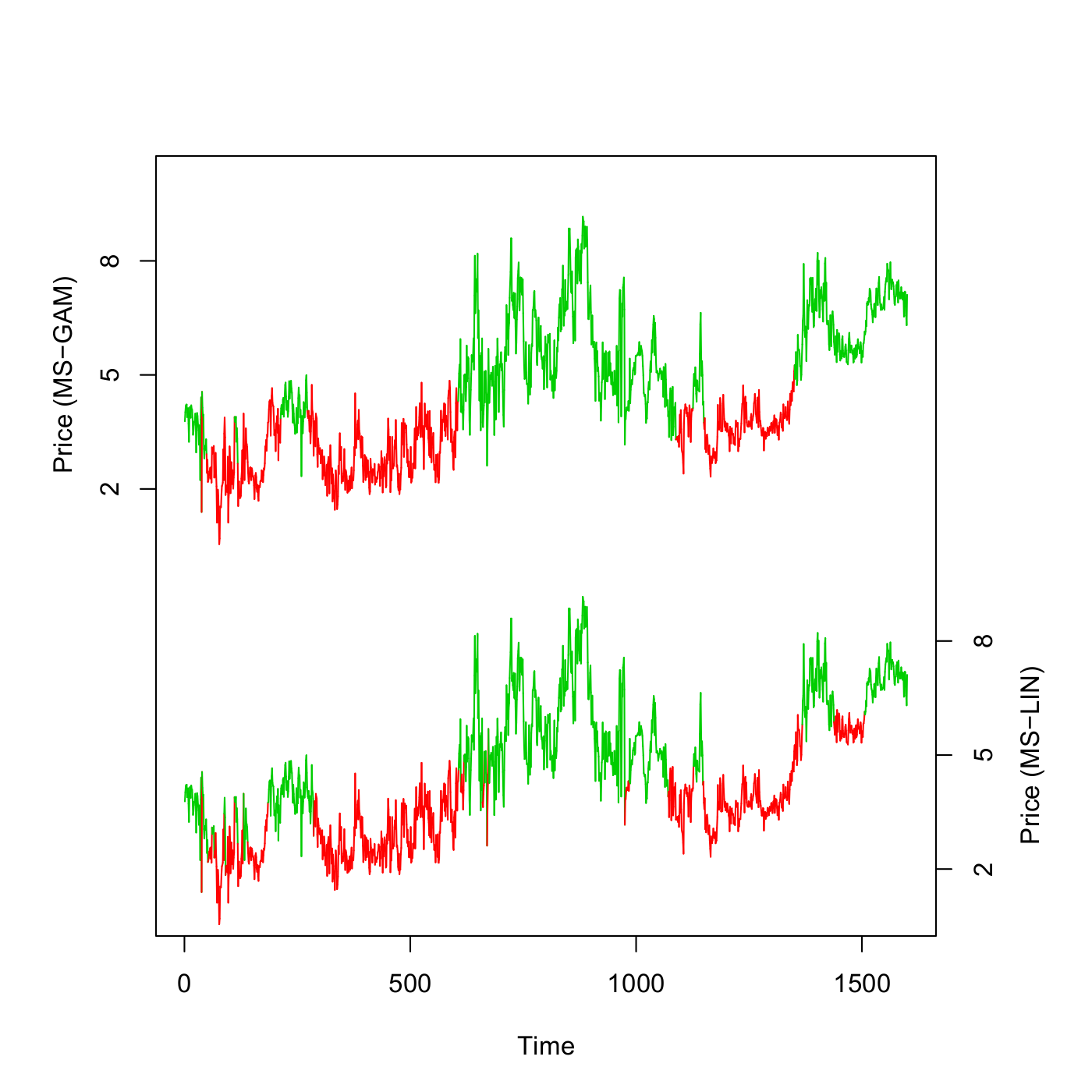}
  \caption{Globally decoded state sequence for the two-state (red and green) MS-LIN and MS-GAM.}
  \label{fig:real_tsseries}
\end{figure}

For the MS-GAM, the transition probabilities were estimated to be $\gamma_{11} = 0.991$ (standard error: $0.006$) and $\gamma_{22} = 0.993$ ($0.003$). The estimated high persistence within states gives evidence that identifiability problems such as those encountered in {\em Scenario III} in the simulation experiments did not occur here. Figure \ref{fig:real_tsseries} gives the estimated regime sequence from the MS-GAM and MS-LIN models obtained using the Viterbi algorithm. Both sequences are similar, with one state relating to occasions where price is more variable and generally higher. However, the MS-LIN model does tend to predict more changes of regime than the MS-GAM, which may be a result of its inflexibility.

\section{Concluding remarks}
\label{discuss}

We have exploited the strengths of the HMM machinery and of penalized B-splines to develop a flexible new class of models, MS-GAMs, which show promise as a useful tool in time series regression analysis. A key strength of the inferential approach is ease of implementation, in particular the ease with which the code, once written for any MS-GAM, can be modified to allow for various model formulations. This makes interactive searches for an optimal model among a suite of candidate formulations practically feasible. Model selection, although not explored in detail in the current work, can be performed along the lines of \citet{cel08} using cross-validated likelihood, or can be based on AIC-type criteria such as the one we considered for smoothing parameter selection. For more complex model formulations, local maxima of the likelihood can become a challenging problem. In this regard, estimation via the EM algorithm, as suggested in \citet{des13} for a smaller class of models, could potentially be more robust (cf.\ \citealp{bul08}), but is technically more challenging, not as straightforward to generalize and hence less user-friendly \citep{mcd14}.

In the example application to energy price data, the MS-GAM clearly outperformed the other models considered, as it accommodates both the need for regime change over time and the need to capture non-linear relationships within a regime. However, even the very flexible MS-GAM has some shortcomings. In particular, it is apparent from the plots, but also from the estimates of the transition probabilities, which indicated a very high persistence of regimes, that the regime-switching model addresses long-term dynamics, but fails to capture the short-term (day-to-day) variations within each regime. In this regard, it would be interesting to explore models that incorporate regime switching (for capturing long-term dynamics induced by persistent market states) but for example also autoregressive error terms within states (for capturing short-term fluctuations). Furthermore, the plots motivate a distributional regression approach, where not only the mean but also variance and potentially other parameters are modeled as functions of the covariates considered. In particular, it is conceptually straightforward to use the suggested type of estimation algorithm also for Markov-switching generalized additive models for location, shape and scale (GAMLSS; \citealp{rig05}).

There are various other ways to modify or extend the approach, in a relatively straightforward manner, in order to enlarge the class of models that can be considered. First, it is of course straightforward to consider semiparametric versions of the model, where some of the functional effects are modeled nonparametrically and others parametrically. Especially for complex models, with high numbers of states and/or high numbers of covariates considered, this can improve numerical stability and decrease the computational burden associated with the smoothing parameter selection. The consideration of interaction terms in the predictor is possible via the use of tensor products of univariate basis functions. The likelihood-based approach also allows for the consideration of more involved dependence structures (e.g.\ semi-Markov state processes; \citealp{lan11}). Finally, in case of multiple time series, random effects can be incorporated into a joint model.

\renewcommand\refname{References}
\makeatletter
\renewcommand\@biblabel[1]{}

\end{spacing}


\markboth{}{}

\begin{thebibliography}{9}

\markboth{}{}


\bibitem[\protect\citeauthoryear{de Boor}{1978}]{deb78}
de Boor, C. (1978),
{\it A practical guide to splines},
Berlin: Springer.

\bibitem[\protect\citeauthoryear{Bulla and Berzel}{2008}]{bul08}
Bulla, J.\ and Berzel, A. (2008),
Computational issues in parameter estimation for stationary hidden Markov models,
\textit{Computational Statistics}, {13}, 1--18.

\bibitem[\protect\citeauthoryear{Celeux and Durand}{2008}]{cel08}
Celeux, G.\ and Durand, J.-P. (2008),
Selecting hidden Markov model state number with cross-validated likelihood,
\textit{Computational Statistics}, {23}, 541--564.

\bibitem[\protect\citeauthoryear{de Souza and Heckman}{2014}]{des13}
de Souza, C.P.E.\ and Heckman, N.E. (2014),
Switching nonparametric regression models,
\textit{Journal of Nonparametric Statistics}, 26, 617--637.

\bibitem[\protect\citeauthoryear{Efron and Tibshirani}{1993}]{efr93}
Efron, B.\ and Tibshirani, R.J. (1993),
{\em An introduction to the bootstrap},
New York, NY: Chapman \& Hall/CRC.

\bibitem[\protect\citeauthoryear{Eilers and Marx}{1996}]{eil96}
Eilers, P.H.C.\ and Marx, B.D. (1996),
Flexible smoothing with $B$-splines and penalties,
\textit{Statistical Science}, {11}, 89--121.


\bibitem[\protect\citeauthoryear{Fr\"uhwirth-Schnatter}{2006}]{fru06}
Fr\"uhwirth-Schnatter, S. (2006),
{\em Finite Mixture and Markov Switching Models},
New York, NY: Springer.

\bibitem[\protect\citeauthoryear{Gassiat {\em et~al.\/}}{in press}]{gas13}
Gassiat, E., Cleynen, A., and Robin, S. (in press).
Inference in finite state space non parametric Hidden Markov Models and applications,
\textit{Statistics and Computing}, DOI:10.1007/s11222-014-9523-8.

\bibitem[\protect\citeauthoryear{Gneiting and Raftery}{2007}]{gne07}
Gneiting, T.\ and Raftery, A.E. (2007),
Strictly proper scoring rules, prediction, and estimation,
\textit{Journal of the American Statistical Association}, {102}, 359--378.

\bibitem[\protect\citeauthoryear{Goldfeld and Quandt}{1973}]{gol73}
Goldfeld, S.M.\ and Quandt, R.E. (1973),
A Markov model for switching regressions,
\textit{Journal of Econometrics}, {1}, 3--16.

\bibitem[\protect\citeauthoryear{Gray}{1992}]{gra92}
Gray, R.J. (1992),
Flexible methods for analyzing survival data using splines, with application to breast cancer prognosis,
\textit{Journal of the American Statistical Association}, {87}, 942--951.

\bibitem[\protect\citeauthoryear{Hamilton}{1989}]{ham89}
Hamilton, J.D. (1989),
A new approach to the economic analysis of nonstationary time series and the business cycle,
\textit{Econometrica}, {57}, 357--384.

\bibitem[\protect\citeauthoryear{Hamilton}{2008}]{ham08}
Hamilton, J.D. (2008),
Regime-switching models,
In: \textit{The New Palgrave Dictionary of Economics}, Durlauf, S.N.\ and Blume, L.E.\ (eds.), Second Edition.

\bibitem[\protect\citeauthoryear{Kim {\em et~al.\/}}{2008}]{kim08}
Kim, C.-J., Piger, J.\ and Startz, R. (2008),
Estimation of Markov regime-switching regression models with endogenous switching,
\textit{Journal of Econometrics}, {143}, 263--273.

\bibitem[\protect\citeauthoryear{Krivobokova {\em et~al.\/}}{2010}]{kri10}
Krivobokova, T., Kneib, T.\ and Claeskens, G. (2010),
Simultaneous confidence bands for penalized spline estimators, 
\textit{Journal of the American Statistical Association}, 105, 852--863.

\bibitem[\protect\citeauthoryear{Langrock {\em et~al.\/}}{in press}]{lan15}
Langrock, R., Kneib, T., Sohn, A.\ and DeRuiter, S.L. (in press),
Nonparametric inference in hidden Markov models using P-splines,
\textit{Biometrics}, DOI:10.1111/biom.12282.

\bibitem[\protect\citeauthoryear{Langrock and Zucchini}{2011}]{lan11}
Langrock, R.\ and Zucchini, W. (2011),
Hidden Markov models with arbitrary state dwell-time distributions,
\textit{Computational Statistics \& Data Analysis}, {55}, 715--724.

\bibitem[\protect\citeauthoryear{MacDonald}{2014}]{mcd14}
MacDonald, I.L. (2014),
Numerical maximisation of likelihood: A neglected alternative to EM?,
\textit{International Statistical Review}, 82, 296--308.


\bibitem[\protect\citeauthoryear{Rigby and Stasinopoulos}{2005}]{rig05}
Rigby, R.A.\ and Stasinopoulos, D.M. (2005),
Generalized additive models for location, scale and shape,
\textit{Journal of the Royal Statistical Society, Series C}, {54}, 507--554.


\bibitem[\protect\citeauthoryear{Sanchez-Espigares and Lopez-Moreno}{2014}]{san14}
Sanchez-Espigares, J.A.\ and Lopez-Moreno, A. (2014),
MSwM: Fitting Markov-switching Models,
R package version 1.2, http://CRAN.R-project.org/package=MSwM.

\bibitem[\protect\citeauthoryear{Wang and Puterman}{2001}]{wan01}
Wang, P.\ and Puterman, M.L. (2001),
Markov Poisson regression models for discrete time series. Part 1: Methodology,
\textit{Journal of Applied Statistics}, {26},  855--869.

\bibitem[\protect\citeauthoryear{Wood}{2006}]{woo06}
Wood, S. (2006),
{\em Generalized Additive Models: An Introduction with R},
Boca Raton, FL: Chapman \& Hall/CRC.


\bibitem[\protect\citeauthoryear{Zucchini and MacDonald}{2009}]{zuc09}
Zucchini, W.\ and MacDonald, I.L. (2009),
{\em Hidden Markov Models for Time Series: An Introduction using R},
Boca Raton, FL: Chapman \& Hall/CRC.

\end{thebibliography}
\end{document}